\documentclass[runningheads]{llncs}
\usepackage{amsmath}
\usepackage{ulem}
\usepackage{multirow}
\usepackage{wrapfig}
\usepackage{verbatim}
 \usepackage[pdftex]{graphicx} \pdfcompresslevel=9

\title{Revisited Experimental Comparison of Node-Link and Matrix Representations}
\author{Mershack Okoe\inst{1} \and Radu Jianu\inst{2} \and Stephen Kobourov\inst{3}}

\authorrunning{Okoe {\it et al.}}
\titlerunning{Revisited Experimental Comparison of Node-Link and Matrix Rep.}

\institute{%
	 Department of Computer Science, Florida International University, Miami, USA\\
	\email{mokoe001@cis.fiu.edu}
\and
giCentre, City, University of London, UK\\
	\email{radu.jianu@city.ac.uk}
	\and 
	Department of Computer Science, University of Arizona, Tucson, USA\\
	\email{kobourov@cs.arizona.edu}
}

\begin{document}

  \maketitle

\begin{abstract}

Visualizing network data is  applicable in domains such as biology, engineering, and social sciences. We report the results of a study comparing the effectiveness of the two primary techniques for showing network data: node-link diagrams and adjacency matrices. Specifically, an evaluation with a large number of online participants revealed statistically significant differences between the two visualizations.
Our work adds to existing research in several ways. First, we explore a broad spectrum of network tasks, many of which had not been previously evaluated. Second, our study uses a large dataset, typical of many real-life networks not explored by previous studies. Third, we leverage crowdsourcing to evaluate many tasks with many participants. 

\end{abstract}

\section{Introduction}

\begin{figure*}[t]
  \centering
  \includegraphics[width=.95\linewidth]{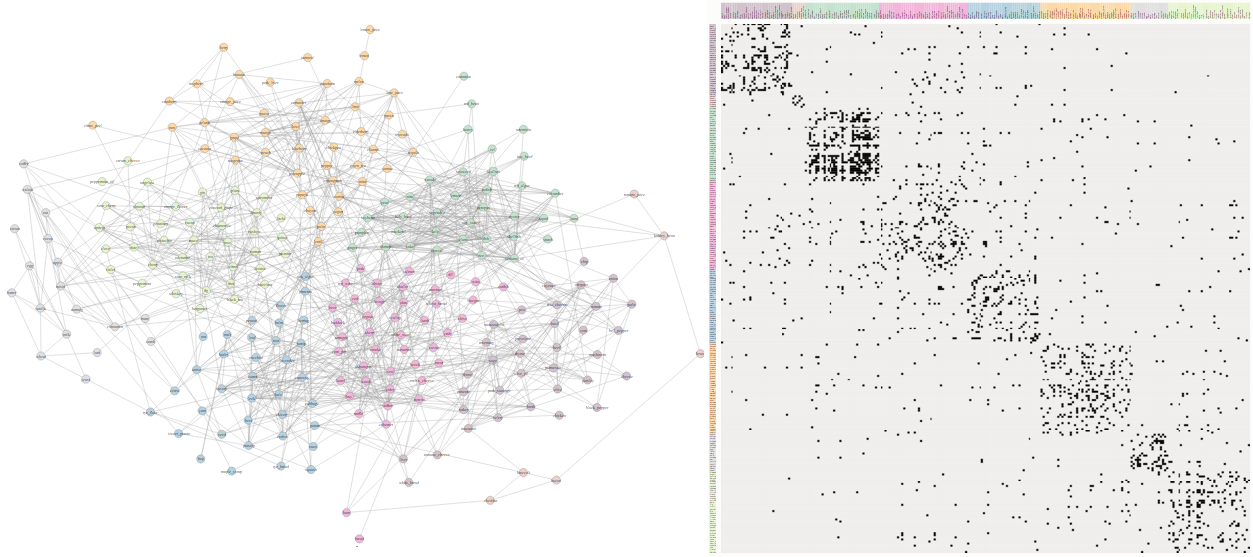}
  \caption{Evaluated visualizations: node-link diagram and adjacency matrix.}
	\label{fig:dataModel}
\end{figure*}

Visualizing network data is known to 
benefit a wide range of domains, including biology, engineering, and social sciences~\cite{von2011visual}. The data visualization community has proposed many approaches to visual network exploration. 
By comparison, the body of work that evaluates the ability of such methods to support data-reading tasks is limited. 
We describe the results of a comparative evaluation of the two most popular ways of visualizing networks: node-link diagrams (NL) and adjacency matrices (AM).   Specifically, we 
consider two interactive visualizations (NL and AM), using a crowdsourced, between-subject methodology, with $557$ distinct online users, $14$ evaluated tasks, and $1$ real-world dataset; see Fig.~\ref{fig:dataModel}.

Several earlier studies compare NL and AM visualizations on specific classes of networks and using a variety of tasks~\cite{ghoniem2004comparison,ghoniem2005readability,okoeecological,keller2006matrices}. They show that the effectiveness of the visualization depends heavily on the properties of the given dataset and the given data-reading tasks. For example, Ghoniem {\it et al.}'s seminal evaluation~\cite{ghoniem2004comparison} found that the two visualizations' ability to support specific tasks depends on the  size and density of the network. Similarly, it is reasonable to hypothesize that there might be differences depending on the   structure of the network (e.g., clustered networks, small-world networks). Thus exploring the effectiveness of NL and AM visualizations on different types of graphs, and using a broader spectrum of tasks, seems worthwhile.

Our study uses one real-world, scale-free dataset of 258 nodes and 1090 edges. This makes our dataset different in structure and larger than previously evaluated networks. For example, Ghoniem {\it et al.} evaluated random networks that were about $2.5$ times smaller, albeit somewhat denser. We argue (in section 3) that our chosen dataset is worth studying as it exemplifies a large class of networks that occur in real applications.

More recently, networks are used to solve increasingly complex problems and as a result, there is an expanding range of tasks that are relevant in real applications and which are of interest to the visualization community. Our study evaluates many tasks ($14$), carefully chosen to span multiple task taxonomies~\cite{lee2006task,amar2005low}. Many of these tasks were not previously investigated in the context of NL and AM representations.

Given the caveat that these results apply to the specific underlying network and the specific implementations of NL and AM visualizations, some of our results confirm prior observations in similar settings, while others are new. 
NL outperforms AM for questions about graph topology (e.g., ``Select all neighbors of node," ``Is a highlighted node connected to a named node?"). 
Of $10$ such tasks, participants who used the node-link diagram were more accurate in $5$ and less accurate in $2$. NL and AM give similar results for $4$ tasks which tested the ability of the participants to identify and compare node groups or clusters,  
except one instance in which AM outperforms NL. Finally, NL and AM provide similar results on $2$ memorability tasks. The full results are shown in Figure 4.

\section{Related Work}

Considerable effort has been expended on optimizing NL and AM visualizations to remove clutter, increase the saliency of visual patterns, and support data reading tasks~\cite{von2011visual}. NL, AM, and slight variations thereof have long been used in practice to support analyses of data in a broad range of domains, including proteomic data~\cite{shannon2003cytoscape,jourdan2003tool,barsky2007cerebral,jianu2014display}, brain connectivity data~\cite{alper2013weighted}, social-networks~\cite{viegas2004social}, and engineering~\cite{sedlmair2011cardiogram}.

Static visual encodings were augmented by interaction to support the exploration and analysis of large and intricate datasets typical of real-life applications. Interactive systems that visualize complex relational data use NL~\cite{auber2004tulip,bastian2009gephi,shannon2003cytoscape}, and AM~\cite{fekete2015reorder,behrisch2014visual,bach2014visualizing,blanch2015dendrogramix,rufiange2012treematrix,bezerianos2010geneaquilts,dinkla2012compressed,sheny2007path}. We reviewed such systems to determine common interactions and included them in our evaluated visualizations.

While the two types of visualizations have been used broadly for a long time, studying how people parse them visually and 
 which visualization method better supports specific tasks and datasets, is ongoing. For example, studies by Purchase {\it et al.}~\cite{purchase1997aesthetic,ware2002cognitive,purchase1996validating} consider how node-link layouts impact data readability, eye-tracking research by Huang {\it et al.} reveal visual patterns and measure the cognitive load associated with network exploration~\cite{huang2007using,huang2009measuring}. More recently Jianu {\it et al.} and Saket {\it et al.} consider the performance of node-link diagrams with overlaid group information~\cite{jianu2014display,saket2014node}.

Our work is one in a series of  studies that compare NL and AM representations. Ghoniem {\it et al.}~\cite{ghoniem2004comparison} evaluated the two approaches on seven connectivity and counting tasks, using interactive visualizations (e.g., node can be selected and highlighted). 
Synthetic graphs of three sizes (20, 50, 100 nodes) and three densities (0.2, 0.4, and 0.6) were used. The authors found that for small sparse graphs, NL was better in connectivity tasks, but that for large and dense graphs, AM outperformed NL for all tasks. Similarly, Keller {\it et al.}~\cite{keller2006matrices} evaluated six tasks on three real-life networks of varying small sizes ($8$, $22$, $50$) and  three densities (unspecified, 0.2, 0.5). Using both static and interactive variants of NL and AM,  Abuthawabeh {\it et al.} found that the participants were equally able to detect structure in graphs representing code dependencies~\cite{abuthawabeh2013finding}. 
Alper {\it et al.} found that in tasks involving the comparison of weighted graphs, matrices outperform node-link diagrams~\cite{alper2013weighted}. Finally, Christensen {\it et al.}~\cite{christensen2014understanding} evaluated matrix quilts in addition to NL and AM in a smaller scale study. 

Our study adds to what is already known in several ways. 
First, we explore a significantly broader range of tasks than earlier studies. These were carefully selected to cover the graph task taxonomy of Lee {\it et al.}~\cite{lee2006task} and the general taxonomy of visualization tasks by Amar {\it et al.}~\cite{amar2005low}. We also considered the task taxonomies for simple graphs~\cite{lee2006task},  clustered graphs~\cite{saket2014group}, and more generally for visualization tasks~\cite{amar2005low,shneiderman1996eyes}, which have been found to be useful in guiding research and informing user study task choices~\cite{jianu2014display,saket2014node}. Second, our study uses a large real-world network, typical of many scale-free networks that arise in practical applications. 
Finally, unlike previous studies, we leverage crowdsourcing, via Amazon's Mechanical Turk, to evaluate many tasks with many participants. 

Note that Mechanical Turk provides access to a diverse participant population~\cite{mason2012conducting,kosara2010mechanical}, and is considered a valid platform for evaluation in general~\cite{paolacci2010running,kosara2010mechanical}, as well as specifically in the context of visualization studies~\cite{heer2010crowdsourcing}. 
Many recent visualization studies 
are crowdsourced~\cite{chapman2014visualizing,micallef2012assessing,jianu2014display,rodgers2015visualizing,borkin2013makes} and specific platforms for online evaluations are developed, including GraphUnit designed for online evaluation of network visualizations~\cite{okoe2015graphunit}.

\section{Study Design}

\subsection{Stimuli: Data}

We evaluated a single network with $258$ nodes and $1090$ edges, representing cooking ingredients connected by edges when frequently used together in recipes. The density of the network was $0.016$ (computed as $\#edges/\#nodes^2$). This network had been explored previously by Ahn {\it et al.}~\cite{ahn2011flavor}. In its original form, the network is larger ($381$ nodes) but we reduced it slightly to ensure it could be visualized smoothly in a browser. We did so by removing disconnected components and low-weight edges.  
Evaluating a single dataset allowed us to cover a broad spectrum of tasks while keeping the size of the study manageable, but naturally, this choice has several limitations, discussed in section 5.

\vspace{2mm}
\noindent\textbf{Rationale:} Our motivation for choosing our network was three-fold. First, it is {\it different than those evaluated already}. Our network is $2.5$ and $5$ times larger than those evaluated by Ghoniem {\it et al.} and Keller {\it et al.}. 
Second, our network was chosen as a {\it representative of several types of real-world networks}. Specifically,  we reviewed $17$ relational datasets (e.g.,  trade exchanges between countries,
the Les Miserable dataset, 
TVCG paper co-authorships, 
protein-interaction networks). We selected one from this set that was representative in terms of structure and density, while at the same time sufficiently small to be evaluated in a browser. Our network has about $4$ times more edges than nodes. This was close to the average edge/node ratio in the $17$ networks we reviewed and 
representative of many networks commonly found in practice~\cite{melancon2006just}. 
Third, we believe a dataset revolving around cooking ingredients would have a {\it greater appeal to participants}. Ingredients were shown as node labels and several tasks referred to ingredients by name. Relatable, concrete dataset may help users understand tasks better~\cite{dagstuhl}.

\subsection{Stimuli: Visual Encoding}

We evaluated two visual encodings: a node-link diagram (NL) drawn using the neato algorithm from graphviz~\cite{graphviz},  and an adjacency matrix (AM), sorted to reveal clusters using the barycenter algorithm available in the Reorder.js library~\cite{fekete2015reorder}. We clustered the network using modularity clustering from GMap~\cite{pacvis10} and encoded this information in the two visual representations using color, as shown in Fig.~1. Both visualizations were developed using the D3 web-library.

\vspace{2mm}
\noindent
\textbf{Rationale:} The neato algorithm is provided in popular layout tools such as graphviz and
frequently part of NL evaluations~\cite{ghoniem2004comparison,jianu2014display}.
We ordered our AM to reveal structure, as we considered this more representative of how matrices are used in practice, unlike  Ghoniem {\it et al.}~\cite{ghoniem2004comparison}, who used a lexicographical order.

\subsection{Stimuli: Interactions}

Both visualizations support panning and zooming, using the mouse-wheel. Multiple nodes can be selected by clicking on them, and deselected with an additional click. Selecting a node in NL colors both the node and its outgoing edges in purple. Selections in AM operate on node labels but change the color of the corresponding node's row or column. Similarly, node mouse-over in NL turns the node and its edges green and shows the node label via tooltips. Node mouse-over in AM colors the row or column. Note that for both node selection and node mouse-over in AM, if a row (column) is colored the complementary column (row) is not.  We chose this approach since both Ghoniem {\it at al.} and Okoe {\it et al.} mention that multiple markings for the same node can confuse users~\cite{ghoniem2004comparison,jianu2014display}.

To select a node as the answer to a task, the participants double-click it. This marks the node with a thick black contour. In both NL and AM this marking was restricted to nodes and labels, without extending to edges or rows/columns. The participants could also deselect an answer by double-clicking it again.

Similar interactions apply to edge selection: An edge mouse-over in NL turns the edge green, and if clicked it is selected and so turns purple. In AM, hovering over an edge-cell highlights its corresponding row and column in green, and clicking it selects the edge.

\vspace{2mm}
\noindent
\textbf{Rationale:} We chose to evaluate interactive visualizations as interactivity is typical in real-world applications. Previous studies, such as those of Ghoniem {\it et al.} or Keller {\it et al.}, also used basic interactions for the same reason. Interactivity can significantly change the effectiveness of a visual encoding, however, and a careful choice of interactive techniques is warranted. 

Our goal was to use interactions that are {\it ecologically valid} (i.e., representative of interactions typical of NL or AM visualizations) and {\it fair} (i.e., providing similar functionality and power in both visualizations). To this end, we reviewed $9$ systems for network visualization (e.g., Gephi~\cite{bastian2009gephi}, Cytoscape~\cite{shannon2003cytoscape}, Tulip~\cite{auber2004tulip}), $12$ network evaluation papers (e.g., Ghoniem {\it et al.}\cite{ghoniem2004comparison}, Keller {\it et al.}~\cite{keller2006matrices}, Okoe {\it et al.}\cite{okoeecological}) 
and $6$ systems and papers for adjacency matrices (e.g., ZAME~\cite{elmqvist2008zame},TimeMatrix~\cite{yi2010timematrix}, work by Perin {\it et al.}~\cite{perin2014revisiting}, work by Henry {\it et al.}~\cite{henry2006matrixexplorer}). We cataloged the interactions described or available in these systems, as well as their particular implementation, and then selected the set of most common interactions.

This resulted in a set of interactions that both overlapped and differed slightly from those implemented in previous studies. Overlapping interactions were described above. New interactions included zooming and panning, which was required to solve some of the tasks. 
We believe the addition of zooming and panning is valuable since such basic navigation is an integral part of real-life systems. Our node-link diagrams also allowed users to move nodes, an interaction that can be used to disambiguate cases in which nodes or edges overlap, and is ubiquitous in NL systems. This interaction does not have an equivalent in AM but is also not necessary as rows and columns are evenly spaced.

\begin{figure}[t]
  \centering
  \includegraphics[width=.43\linewidth]{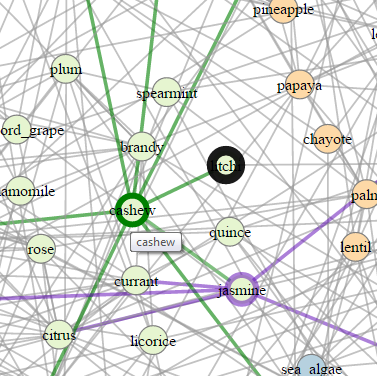}\hspace{.5cm} \includegraphics[width=.43\linewidth]{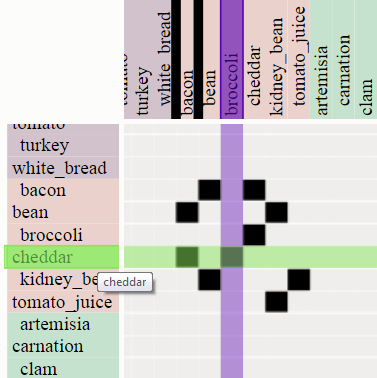}
  \caption{Participants mouse-over nodes to highlight them (green) and click on nodes to select them (purple). Designating a node as the answer for a task answer is accomplished via a double-click, which draws a black contour around the node.}
	\label{fig:interactions}
\end{figure}

\subsection{Tasks}

We evaluated the $14$ tasks described in Table 1. Participants solved multiple repeats (generally $5$ or $10$) of each task. Task repeats were selected manually on the network so as to cover multiple levels of complexity. For example, our repeats included nodes with both low and large degrees (e.g., $T1$, $T2$), short and long paths (e.g., $T10$, $T13$), or nodes with few and many neighbors (e.g., $T4$). 

Three of our tasks warrant a more detailed discussion. We included two memorability tasks, ($T11$, $T14$). The former tested the ability of participants to recall data they had looked for or accessed at an earlier time, and is similar to memorability tasks evaluated by Saket {\it et al.}~\cite{Memorability_Saket2015}. The latter tested the ability of participants to recognize visual configurations they had viewed previously and is more similar to tasks used by Jianu {\it et al.} and Borkin {\it et al.}~\cite{jianu2014display,borkin2013makes}. Both memorability tasks were based on questions that the participants had to answer early in their session (i.e., $T9$ in group 4, and $T12$ in group 5) to prime the participants with a particular piece of information or visual configuration. A few minutes later, after performing a set of other  tasks (i.e, $T10$ in group 4, $T13$ in group 5), the participants were asked about the information from the earlier task. Finally, we added a path-estimation task ($T5$), which required the participants to estimate how far two nodes are, in terms of the shortest path between them. Timing constraints ensured that participants used perceptual mechanisms to give a best-guess response instead of ``computing'' the correct answer.  

\vspace{2mm}
\noindent
\textbf{Rationale:} Our overarching goal in selecting our tasks was to cover a wide spectrum of different and realistic network tasks. We selected tasks to cover the graph objects they provide answers about (i.e., nodes, edges, paths), as well as cover Lee {\it et al.}'s categories of graph-reading tasks, and Amar {\it et al.}'s general types of visualization tasks. Several of our tasks have been used before but under slightly different conditions. Additionally, 
we included tasks that go beyond previous studies comparing NL and AM, such as tasks involving clusters. 
We also included memorability tasks as they are a topic of growing interest in the visualization community~\cite{borkin2013makes,Memorability_Saket2015}. We also hypothesized there would be differences between the two visualizations in this respect. 
We included a path-estimation task~\cite{jianu2014display}, as it is a good representative of the ``Overview'' category of graph tasks, and underlies perceptual queries that users make on relational data.

\begin{table*}[t]						{\tiny
	\centering					
		\begin{tabular}{|l|l|l|l|l|l|l|}				
		\hline			
Task&	Target&	Task tax.~\cite{lee2006task}&	Task tax.~\cite{amar2005low}&	Group&	\#Repeats&	Time\\\hline
\shortstack[l]{1. Given two highlighted nodes, select the \\one with the larger degree.}&	\shortstack[l]{node}&	\shortstack[l]{Topology\\ (adjacency)}&	\shortstack[l]{Retrieve value,\\ Sort}&	\shortstack[l]{1}&	\shortstack[l]{10}&	\shortstack[l]{15}\\\hline
\shortstack[l]{2. Given a highlighted node, select all its \\neighbors}&	\shortstack[l]{edge}&	\shortstack[l]{Topology \\(adjacency, \\accessability)}&	\shortstack[l]{Retrieve value, \\Filter}&	\shortstack[l]{1}&	\shortstack[l]{10}&	\shortstack[l]{25}\\\hline
\shortstack[l]{3. Given two clusters of highlighted nodes, \\which one is more interconnected?}&	\shortstack[l]{clusters,\\ cliques}&	\shortstack[l]{Overview \\(connectivity)}&	\shortstack[l]{Filter, Sort, \\Cluster}&	\shortstack[l]{1}&	\shortstack[l]{10}&	\shortstack[l]{10}\\\hline
\shortstack[l]{4. Given two highlighted nodes, select all \\of the common neighbors.}&	\shortstack[l]{edge}&	\shortstack[l]{Topology\\ (shared\\ neighbor)}&	\shortstack[l]{Retrieve value,\\ Filter}&	\shortstack[l]{2}&	\shortstack[l]{10}&	\shortstack[l]{30}\\\hline
\shortstack[l]{5. Given two pairs of highlighted nodes \\(red and blue) and limited time, estimate\\ which pair is closer in terms of graph \\topology?}&	\shortstack[l]{path, \\edge}&	\shortstack[l]{Overview\\ (connectivity)}&	\shortstack[l]{Derive value,\\ Sort}&	\shortstack[l]{2}&	\shortstack[l]{10}&	\shortstack[l]{10}\\\hline
\shortstack[l]{6. How many clusters are there in the\\ visualization? \\ $^\ast$clusters shown via color (section 3.2)}&	\shortstack[l]{clusters}&	\shortstack[l]{Overview\\ (connectivity)}&	\shortstack[l]{Derive \\value}&	\shortstack[l]{3}&	\shortstack[l]{1}&	\shortstack[l]{10}\\\hline
\shortstack[l]{7. Given two groups of highlighted nodes \\(e.g., red and blue) and limited time, \\estimate which group is larger. }&	\shortstack[l]{clusters}&	\shortstack[l]{Attribute \\based}&	\shortstack[l]{Filter, Sort,\\ Derive value, \\Correlate}&	\shortstack[l]{3}&	\shortstack[l]{10}&	\shortstack[l]{10}\\\hline
\shortstack[l]{8. Given two highlighted nodes decide \\whether they belong to the same cluster. \\ $^\ast$clusters shown via color (section 3.2)}&	\shortstack[l]{clusters,\\ nodes}&	\shortstack[l]{Attribute\\ based}&	\shortstack[l]{Cluster, \\Filter}&	\shortstack[l]{3}&	\shortstack[l]{10}&	\shortstack[l]{10}\\\hline
\shortstack[l]{9. Given one highlighted node and one \\named node, are they connected?}&	\shortstack[l]{edge}&	\shortstack[l]{Topology\\ (adjacency)}&	\shortstack[l]{Retrieve value}&	\shortstack[l]{4}&	\shortstack[l]{5}&	\shortstack[l]{20}\\\hline
\shortstack[l]{10. Given two highlighted nodes, how long \\is the shortest path between them?}&	\shortstack[l]{path, \\edge}&	\shortstack[l]{Topology\\ (connectivity)}&	\shortstack[l]{Retrieve value, \\Derived value,\\ filter}&	\shortstack[l]{4}&	\shortstack[l]{5}&	\shortstack[l]{60}\\\hline
\shortstack[l]{11. Memorability: After spending several \\minutes on task 10, can participants\\ remember the answers they gave to \\task 9, without access to the visualization?}&	\shortstack[l]{}&	\shortstack[l]{See section 3.4}&	\shortstack[l]{See section 3.4}&	\shortstack[l]{4}&	\shortstack[l]{5}&	\shortstack[l]{unlim}\\\hline
\shortstack[l]{12. Given two highlighted nodes and three \\named ones, which of the named nodes \\is connected to both highlighted nodes? \\(exemplified in Figure 3)}&	\shortstack[l]{edge}&	\shortstack[l]{Topology \\(shared neighbor)}&	\shortstack[l]{Retrieve value,\\ Filter}&	\shortstack[l]{5}&	\shortstack[l]{5}&	\shortstack[l]{60}\\\hline
\shortstack[l]{13. Given a selected node, how many nodes \\are within two edges' reach?}&	\shortstack[l]{edge}&	\shortstack[l]{Topology\\ (accessibility)}&	\shortstack[l]{Retrieve value,\\ Derive value, \\Filter}&	\shortstack[l]{5}&	\shortstack[l]{5}&	\shortstack[l]{60}\\\hline
\shortstack[l]{14. Memorability: After spending several \\minutes on tasks 13, can participants remember \\(i.e., select) which nodes were highlighted as \\part of task 12, if showed the visualization \\with the answers they gave to task 13 \\highlighted?}&	\shortstack[l]{}&	\shortstack[l]{**See paper \\body}&	\shortstack[l]{**See paper \\body}&	\shortstack[l]{5}&	\shortstack[l]{5}&	\shortstack[l]{unlim}\\\hline
\end{tabular}	
\vspace{.1cm}
		\caption{Tasks: the columns describe (i) the task, (ii) targeted network element, (iii-iv) task categories in Lee {\it et al.}'s and Amar {\it et al.}'s taxonomies, (v) group number the task was evaluated in, (vi) number of instances of this task type, (vii) task time limit (sec).	}			
		\label{tab:Table1}				}
\end{table*}

\subsection{Hypotheses}

Based on previous results by Ghoniem {\it et al.}~\cite{ghoniem2004comparison}, Keller {\it et al.}~\cite{keller2006matrices}, Okoe {\it et al.}~\cite{okoe2015graphunit}, Jianu {\it et al.}~\cite{jianu2014display}, and Saket {\it et al.}~\cite{saket2014node} we devised the null hypotheses:

\begin{itemize}
\item[] \textbf{H1}: There is no statistically significant difference in time and accuracy performance between using NL and AM for tasks involving the retrieval of information about nodes and direct connectivity ($T1$, $T2$, $T4$, $T9$, $T12$). 

\item[] \textbf{H2}: There is no statistically significant difference in time and accuracy performance between using NL and AM for connectivity and accessibility tasks involving paths of length greater than two ($T5$, $T10$, $T13$). 

\item[] \textbf{H3}: There is no statistically significant difference in time and accuracy performance between using NL and AM on group tasks ($T3,T6,T7,T8$).

\item[] \textbf{H4}: There is no statistically significant difference in memorability between using NL and AM.

\end{itemize}

\noindent We expected H1 to hold and H2 not to hold. We also thought H3 would hold, except for estimating group interconnectivity ($T6$), since estimating the number of non-overlapping dots in a square (AM) should be easier than estimating overlapping edges in an irregular 2D area (NL). Finally, we anticipated memorability would be higher in node-link diagrams due to its more distinguishable features.

\subsection{Design}

We used a between-subjects experiment with two conditions. We divided our $14$ task types into $5$ experimental groups, as shown in Table 1, and we evaluated each group separately. Each participant was allowed to participate in a single group and used just one of the two visualizations. We assigned participants to groups and conditions in a round-robin fashion. We aimed to collect data from around $50$ participants per condition. As  some participants did not complete the study, the total number of participants for whom we collected data varies slightly between conditions. 
All tasks were timed as shown in Table 1, with time limits determined experimentally through a pilot-study and chosen to allow most participants to complete the tasks, while moving the study along.  

\vspace{2mm}
\noindent
\textbf{Rationale}: Between-subject experiments are frequently used in the visualization community~\cite{jianu2014display,saket2014node,ziemkiewicz2008shaping,borkin2011evaluation,robertson2008effectiveness,kosara2010mechanical,micallef2012assessing}. One advantage of this design is the absence of learning effects between evaluated conditions. A disadvantage is the need for large numbers of participants, which is easily mitigated in a crowdsourced setting. Moreover, between-subjects designs are quicker (since only one condition is evaluated at a time) and  online participants prefer shorter studies.

We divided the tasks into groups for the same reason. Having each participant evaluate all tasks 
would have resulted in excessively long sessions that participants would have found tiring. Having participants solve only subsets of tasks allowed us to reduce their time commitment. 
We used estimated task completion times to group tasks, aiming for an expected duration of about $15$ minutes.

We aimed for $50$  participants per condition, matching the numbers used in earlier crowdsourced studies~\cite{chapman2014visualizing,jianu2014display}. We decided to enforce short time-limits in order to limit and make uniform the total session duration across participants.

\subsection{Procedure}

We used Amazon's Mechanical Turk (MTurk) to crowdsource our study to a broad population. To account for variations in participant demographics during the day, we published study batches throughout the day. We ran conditions in parallel and directed incoming participants to them using a round-robin assignment, to ensure that the two conditions sampled participants from the same populations. The demographics of MTurk users are reported by Ross {\it et al.}~\cite{ross2010crowdworkers}.

Each incoming participant was first presented with an introduction to the study, dataset, the visualization they would see and use, and the tasks they would perform. Each task was exemplified in the introduction, as shown in Figure 2. Since our interactions relied on color, participants were administered a color-blindness test. Next came a training session which involved solving two instances of each type of task in their assigned group. During the training session the participants could check the correctness of their answers.

Finally, the participants were lead to the main part of the study. 
In the main part of the study, task instances of each type in an assigned group were shown to the participants. 
For example, since group $1$ involved three distinct task types, participants assigned to it solved three consecutive sections of ten task-instances each. 
At the end, we asked the participants for comments.

We used GraphUnit~\cite{okoe2015graphunit} to create the study, deploy it, and collect data. Visualizations were shown on the left, while task instructions and answer widgets were shown on the right. Depending on each task, users answered by selecting nodes or by using interactive widgets (e.g., text-boxes, check-boxes). Time limits were enforced by showing  a count-down timer and hiding the visualization once the counter expired. To increase the chances of collecting clean data we awarded a bonus to the best result in each group and told participants that two of the task-instances were control tasks easy enough for anyone to solve.


\section{Results}

Our results are summarized in Fig.~\ref{fig:results}. By and large, they show that node-link diagrams were better for most types of connectivity tasks ($T1$, $T2$, $T4$, $T5$, $T9$, $T10$, $T13$) thereby invalidating both  H1 and H2. The fact that H1 does not hold is surprising given previous results. Performance on group tasks was generally comparable with the two visualizations, as hypothesized ($H3$), though we found that the AM was better for estimating the number of clusters rather than their interconnectivity. Finally, NL supported memorability tasks better (invalidating $H4$). In particular, NL users outperformed AM users when recalling previously used data ($T11$). 

\subsubsection{Data processing:}

\begin{figure*}[t]
  \centering
  \includegraphics[width=0.44\linewidth]{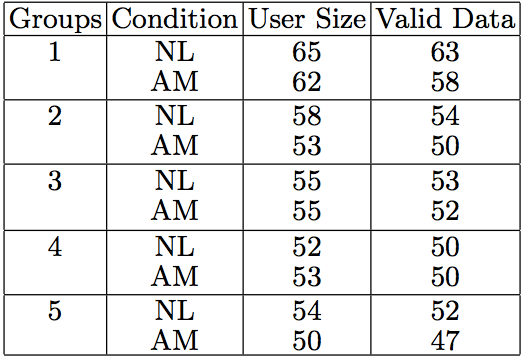}
  \caption{Number of participants in each task group per condition and the number of valid submissions used after data cleaning.}
	\label{tab:groups_and_users}
\end{figure*}

We collected data from 557 individual participants distributed across task groups and conditions as shown in Fig.~\ref{tab:groups_and_users}. We removed a total of $28$ responses from participants who spent an average of $2$ seconds per task and had accuracy in the bottom $10$ percentile. We considered these likely to be random responses by participants attempting to game the study.

To compute the accuracy of node selections ($T1$, $T2$, $T4$), we used the formula $Acc = (\|PS \cap TA\|)/\|TA\|\}$, where $PS$ is the participant's selection and $TA$ is the true answer. To compute answers for tasks involving numeric answers ($T6$, $T10$, $T13$) we used the formula $Acc=max(0,1-\|PA-TA|/|TA|)$, where $PA$ is the participant's answer and $TA$ is the true answer. For other tasks we gave a $1$ to correct answers, and a $0$ to incorrect answers. Since each task type was represented in the study by several repeats, we averaged the accuracies of a task's individual repeats into an accuracy for the task as a whole. 

\subsubsection{Statistical analysis:}

\begin{figure*}[t]
  \centering
  \includegraphics[width=0.95\linewidth]{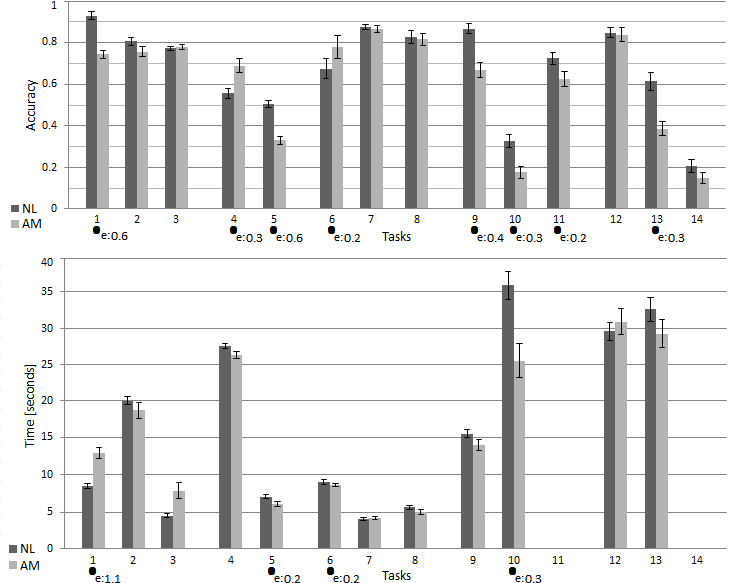}
  \caption{Results: accuracy and time. Error bars show one standard error. Statistically significant results and effect sizes are also marked. Tasks $14$, $11$ had no time limits.}
	\label{fig:results}
\end{figure*}

If the data is normally distributed (determined via a Shapiro-Wilk test) we use a t-test analysis between conditions to determine if the observed differences are significant. Otherwise we use a Wilcoxon-Rank-Sum test. We indicate statistically significant differences and effect sizes in Fig.~\ref{fig:results}.

\section{Discussion}

Based on the quantitative results and our own interactions with the visualizations, we believe the results can be explained by several factors.

First, NL can be more compact than AM since their layout fully leverages the 2D area, while matrices are constrained to two 1D linear node orders. Matrices favor dense networks (as number of edges increases, matrix size remains constant) but not sparse ones (empty matrices are as large as a dense ones). Instead, sparse NL diagrams can be packed tightly. At the extreme, an empty network can be shown without loss in readability using NL in a $\sqrt{N}\times\sqrt(N)$ square. The same empty network would require a $N \times N$ square in an AM.
Thus, as networks grow larger but not necessarily denser, AM may incur an increasing navigation cost. Concretely, our NL diagrams required less zooming for nodes to become legible and selected accurately. This could explain the differences in $T1$.

Second, NL draw a node's glyph and connections together. Thus, once a label is spotted, from it, its outgoing edges can be traced to other nodes and their labels. Moreover, the presence of the edge aids this tracing.  Instead, matrices show node information and edge information separately. Finding the endpoints of an edge involves two potentially long visual-traces along the horizontal and vertical axes. Similarly, finding an edge of an identified node involves a horizontal or vertical search. This could be one of the reasons for the large effect in $T9$. However,  this described behavior is only hypothesized and yet to be demonstrated.

Third,  Ghoniem {\it et al.} found that AM performs poorly on tasks involving long paths~\cite{ghoniem2004comparison}, and our results on $T10$ and $T13$ confirm this. Interestingly, the average time of participants performing path tasks ($T10$) in AM is significantly shorter than that for NL. However, we found that this is due to many AM users giving up on solving the task altogether early on. Moreover, NL layouts aim to place nodes so that their network distance matches their embedded distance. While matrices can also order rows and columns, they are constrained by the use of a single dimension. This could explain the results of $T5$: when one pair of nodes were in the same cluster and the other not, comparing their topological proximity was possible in both visualizations, but in all other cases NL outperforms AM.

Matrices eliminate occlusion and ambiguity problems. In NL diagrams it is sometimes difficult to tell if an edge connects to a node or passes through it, but this is not the case in AMs.  Moreover, many tasks that involve visual searches in unconstrained 2D space with NL, are easier with AM.
For example, finding a node in an AM involves a linear scan in 
a list of labels. Counting nodes with certain properties can also be done sequentially by moving through the matrix's headers. Such tasks are difficult in NL diagrams as users have to search a 2D space and keep track of already visited nodes.  This may account for $T4$, where AM outperforms NL: participants could systematically scan two selected AM node-rows and identify the columns where both rows had an edge. 

\subsubsection{Limitations:}

Several earlier studies comparing NL and AM considered the effects of network size and density~\cite{ghoniem2005readability,keller2006matrices}. While we recognize the value of this approach, this was beyond the scope of our current study. Instead, we aimed to understand how the two visualizations support a more complete range of tasks (14 versus previously 7 and 6) in a network that is representative of real-world networks in size and structure. It is unclear whether our results would generalize to real-world networks that are significantly larger or denser but our work does provide additional experimental data for a network unlike those evaluated earlier. 

We use one type of network and a single instance thereof. This is a methodological drawback which we accepted, due to the overhead associated with preparing multiple appropriate real-world networks for evaluation and phrasing participant instructions using the semantics of different networks. 
While the limitations of this approach are non-trivial, we attempted to balance them by using multiple task-repeats of the same type and focusing on different parts of the network.

The density of our network was significantly lower than~\cite{ghoniem2004comparison,keller2006matrices}.
However, Melancon points out that large real-world networks with high densities are rare~\cite{melancon2006just}. He argues that the edge-to-node ratio is a better indicator for density in real-world networks as it is less sensitive to the number of nodes. Indeed, only $1$ of the $17$ networks we considered, and $3$ of the $19$ networks Melancon considered had densities higher than $0.2$. In $3$ of these $4$ cases, these dense networks were also the smallest in terms of number of nodes.   

As in recent studies, we  evaluate interactive visualizations. Given the  different visual encoding in NL and AM it is difficult to ensure that all interactions are fair to both visualizations. To alleviate this concern 
we relied on a detailed review of the NL and AM literature, and selected the most common interactions and their implementations (see Section 3.3). This ensured, at least to some degree, that we evaluated the interactive visualizations as they appear in practice.

Crowdsourced studies have known inherent limitations (e.g., difficulty controlling the experimental setup and verifying what participants do). 
By and large, however, crowdsourcing studies replicate prior controlled lab studies~\cite{heer2010crowdsourcing}.

\section{Conclusions}

We presented the results of a crowdsourced evaluation of NL and AM network visualizations. Our study involved 557 online participants who used interactive versions of the two encodings, to answer $14$ varied types of questions about a large network of $256$ nodes and $1090$ edges. We found that NL is better than AM for questions about network topology and connectivity, and comparable for group and memorability tasks, and therefore a better choice
for visualizing datasets similar to the one we evaluated, provided a similar interaction set.

\newpage
\bibliographystyle{splncs03}
\bibliography{bib}

\end{document}